\documentclass[twocolumn,twoside,slac_two]{revtex4}
\usepackage{graphicx}
\usepackage{fancyhdr}
\newcommand{\lsim}{{\lower.5ex\hbox{$\; \buildrel < \over \sim \;$}}}
\newcommand{\gsim}{{\lower.5ex\hbox{$\; \buildrel > \over \sim \;$}}}
\newcommand{\fermi}{{\it Fermi} }
\newcommand{\planck}{{\it Planck}}
\newcommand{\swift}{{\it Swift} }
\def\nup{$\nu_{\rm peak}$}
\def\ergs{{erg~cm$^{-2}$~s$^{-1}$~}}
\def\avvovm{{$\langle V/V_{\rm m} \rangle$}} 
\def\gr{{$\gamma$-ray~}}

\begin{document}

\title{A simplified view of blazars: why BL Lacertae is actually a quasar
  in disguise}

%

\author{P. Padovani}
\affiliation{European Southern Observatory, Karl-Schwarzschild-Str. 2,
  D-85748 Garching bei M\"unchen, Germany}
\author{P. Giommi, G. Polenta$^a$, S. Turriziani, V. D'Elia$^a$}
\affiliation{ASI Science Data Center, c/o ESRIN, via G. Galilei, I-00044
  Frascati, Italy}
\author{S. Piranomonte}
\affiliation{$^a$INAF-Osservatorio Astronomico di Roma, via Frascati 33, I-00040
Monteporzio Catone, Italy}

\begin{abstract}
We put forward a scenario where blazars are classified as flat-spectrum
radio quasars, BL Lacs, low synchrotron, or high synchrotron peaked
objects according to a varying combination of Doppler boosted radiation
from the jet, emission from the accretion disk, the broad line region, and
light from the host galaxy. We thoroughly test this new approach, which
builds upon unified schemes, using Monte Carlo simulations and show that it
can provide simple answers to a number of long-standing open issues. We
also demonstrate that selection effects play a very important role in the
diversity observed in radio and X-ray samples and in the correlation
between luminosity and the peak frequency of the synchrotron power (the so-called ``blazar
sequence''). It turns out that sources so far classified as BL Lacs on the
basis of their observed weak, or undetectable, emission lines are of two
physically different classes: intrinsically weak-lined objects, more common
in X-ray selected samples, and heavily diluted broad-lined sources, more
frequent in radio selected samples, which explains some of the confusion in
the literature.
\end{abstract}

\maketitle

\thispagestyle{fancy}


\section{INTRODUCTION}
Blazars are radio loud active galactic nuclei (AGN) pointing their jets in
the direction of the observer \cite{bla78,UP95}. While all blazars emit
variable, non-thermal radiation across the entire electromagnetic spectrum,
they come in two main subclasses, whose major difference is in their
optical properties: 1) Flat Spectrum Radio Quasars (FSRQs), which show
strong, broad emission lines in their optical spectrum, just like radio
quiet quasars; and 2) BL Lacertae objects (BL Lacs), which are instead
characterized by an optical spectrum, which at most shows weak emission
lines, sometimes displays absorption features, and in some cases can be
completely featureless. Historically, the separation between BL Lacs and
FSRQs has been made at the (rather arbitrary) rest-frame equivalent width
(EW) of 5~\AA~\cite{stickel1991,sto91}. However, no evidence for a bimodal
distribution in the EW of the broad lines of radio quasars has ever been
found and, on the contrary, radio-selected BL Lacs were shown to be, from
the point of view of the emission line properties, very similar to FSRQs
but with a stronger continuum \cite{sca97}. Most BL Lacs selected in the
X-ray band, on the other hand, had very weak, if any, emission lines, and,
when studying the properties of the X-ray selected {\it Einstein} Medium
Sensitivity Survey (EMSS) sample, another criterion to identify them, this
time to separate them from galaxies, had to be introduced
\cite{sto91}. This was based on the Ca H\&K break, a stellar absorption
feature typically found in the spectra of elliptical galaxies. Given that
its value in non-active ellipticals is $\sim 50\%$, a maximum value of
$25\%$ was chosen to ensure the presence of a substantial non-thermal
continuum superposed to the host galaxy spectrum \cite{sto91}. This was
later revised to $40\%$ \cite{mar96,lan02}.

\begin{figure*}[t]
\includegraphics[height=7.8cm,angle=-90]{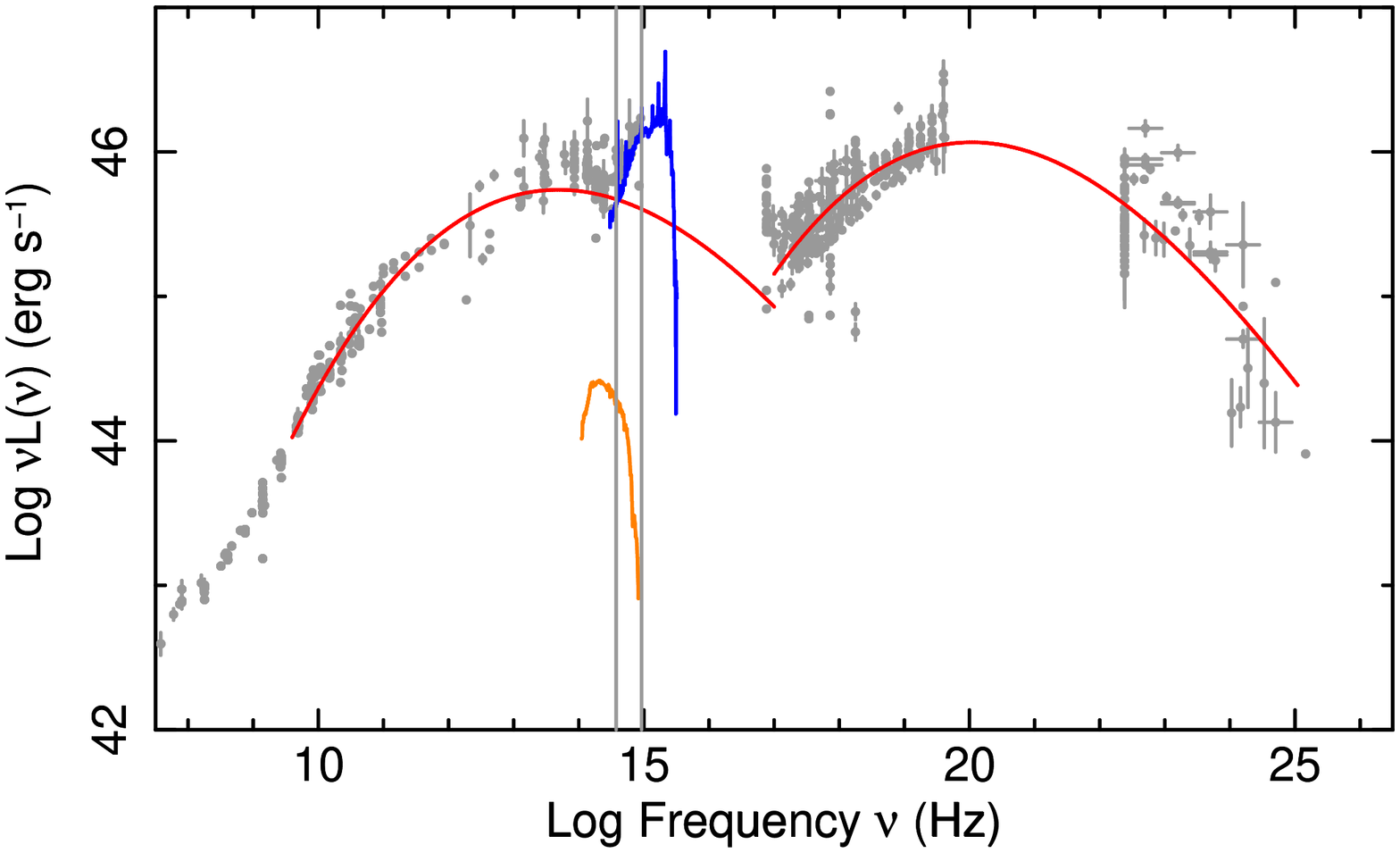}
\hspace{0.3cm}
\includegraphics[height=7.8cm,angle=-90]{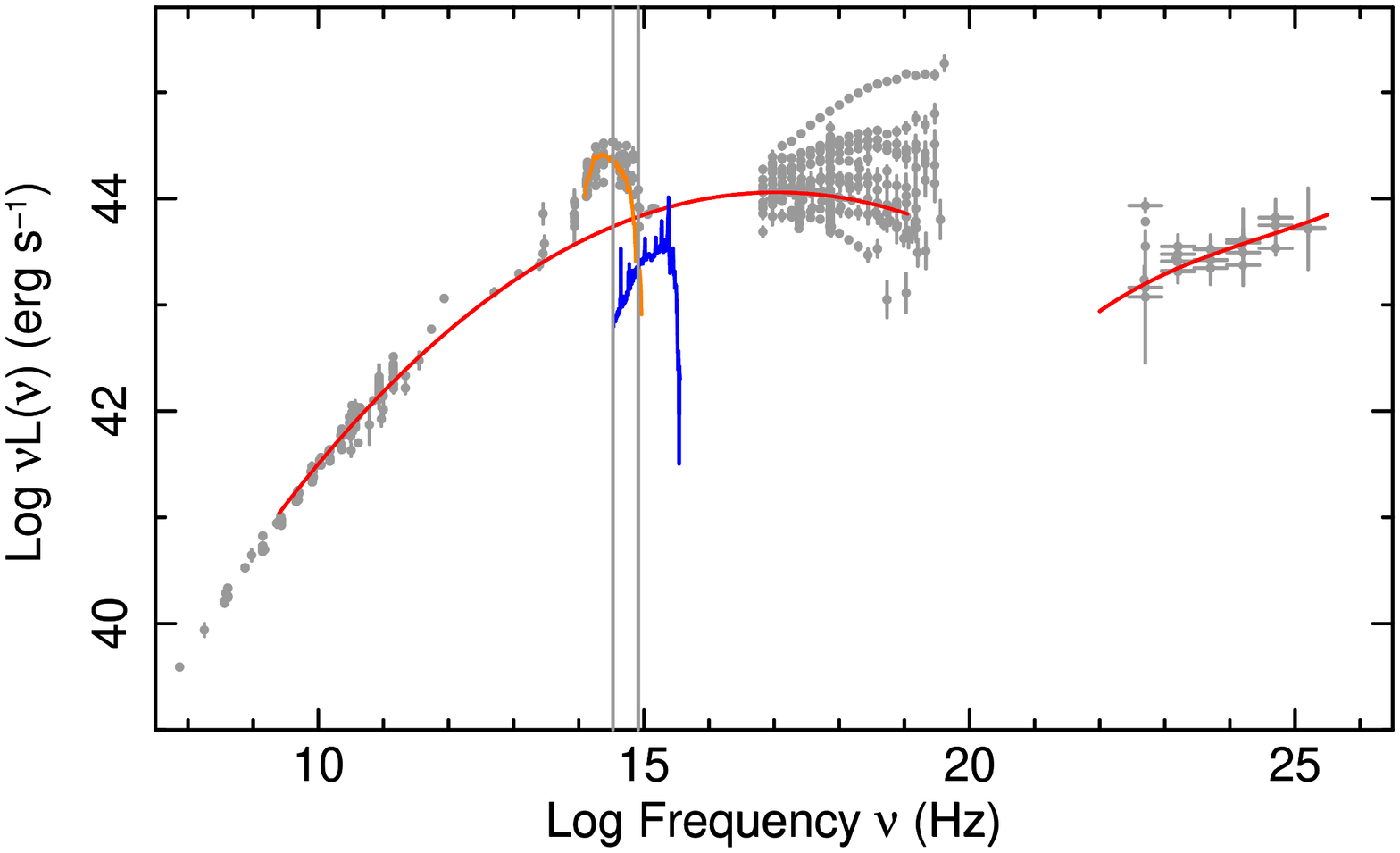}\\
\vspace{2.5cm}
\includegraphics[height=7.8cm,angle=-90]{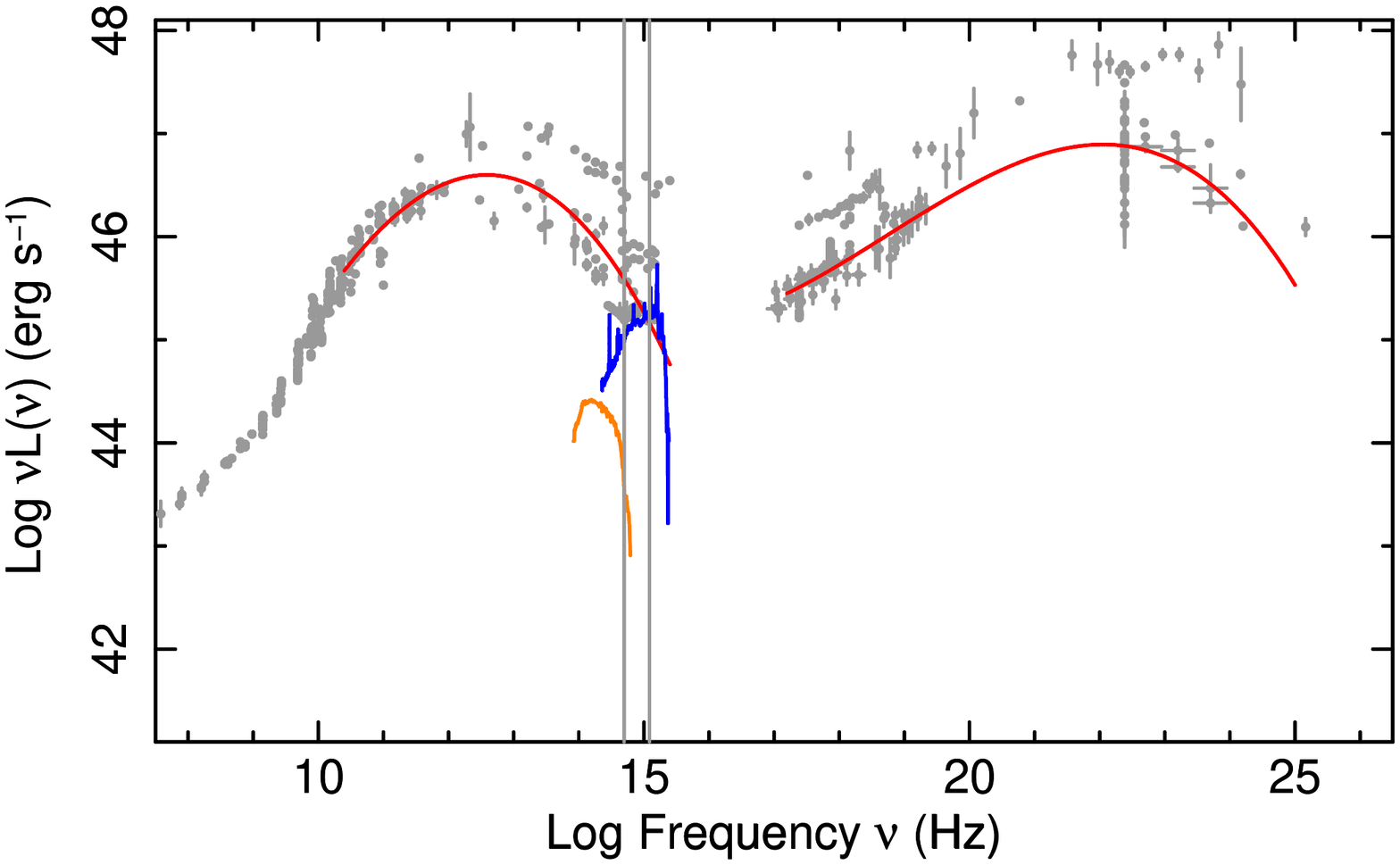}
\hspace{0.3cm}
\includegraphics[height=7.8cm,angle=-90]{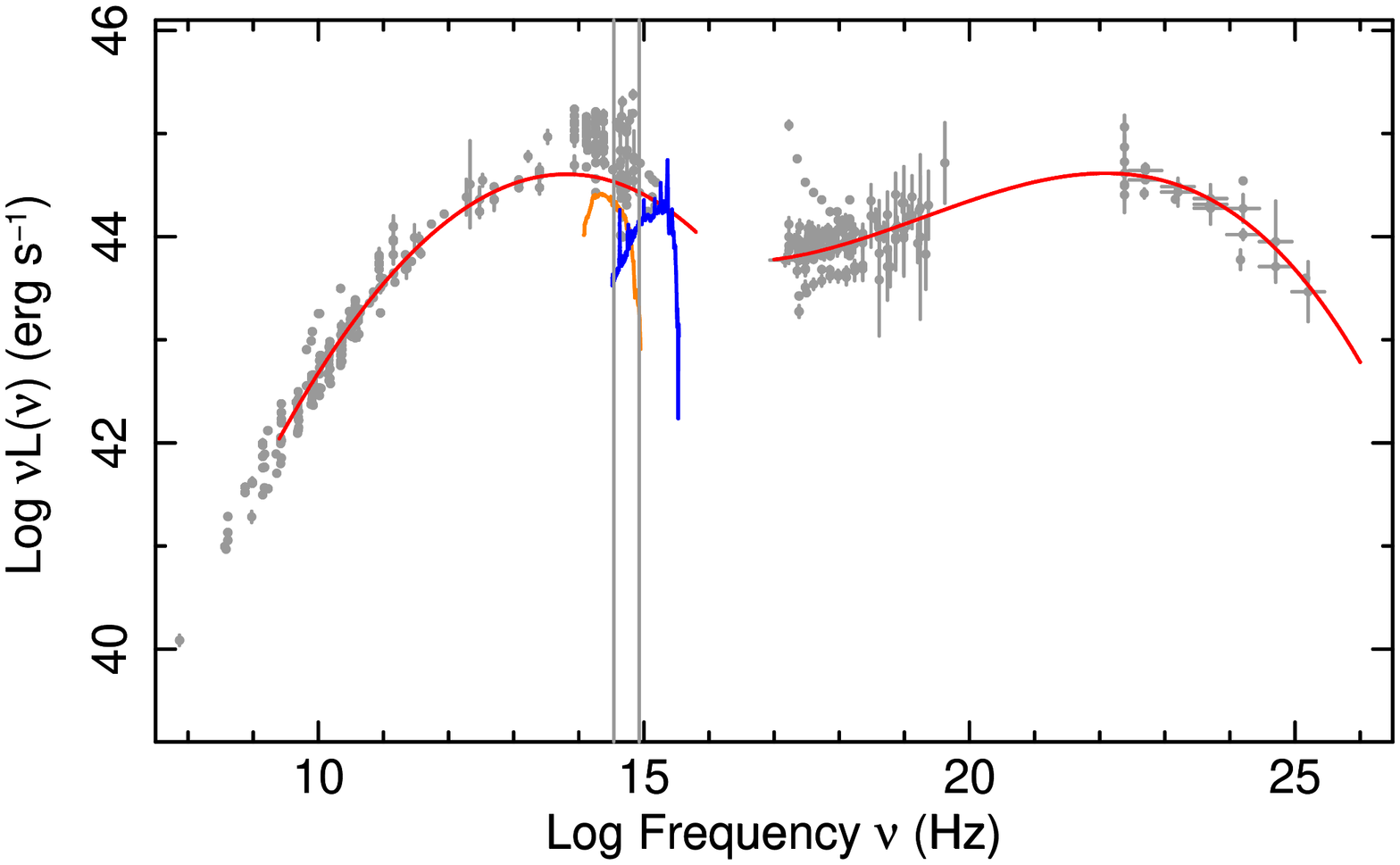}
\vspace{2.0cm}
\caption{The SEDs of four representative blazars: two FSRQs, 3C 273 and 3C
  279, and two BL Lacs, MKN 501 and BL Lac. The lines in color denote the
  three main components of blazars SEDs, namely non-thermal radiation from
  the jet (red), emission from the disk and from the broad line region
  represented by the composite QSO optical spectrum of \protect
  \cite{van01} (blue), and light from the host galaxy, represented by the
  giant elliptical template of \protect \cite{man01} (orange).  The two
  vertical lines indicate the optical observing window ($3800 - 8000$
  \AA).}
\label{fig:sed}
\end{figure*}

Blazar classification depends then on the details of their appearance in
the optical band where they emit a mix of three types of radiation: 1) a
non-thermal, jet-related, component; 2) thermal radiation coming from the
accretion onto the supermassive black hole and from the broad line region
(at least in most radio-selected sources); 3) light from the host (giant
elliptical) galaxy. Figure \ref{fig:sed} represents these three components
as red, blue and orange lines, overlaid to the spectral energy distribution
(SED) of four well-known blazars \cite{GiommiPlanck}. The strong
non-thermal radiation, the only one that spans the entire electromagnetic
spectrum, is composed of two basic parts forming two broad humps, the
low-energy one attributed to synchrotron radiation, and the high-energy
one, usually thought to be due to inverse Compton radiation
\cite[e.g.][]{abdosed}.  The peak of the synchrotron hump (\nup) can occur
at different frequencies, ranging from about $\sim 3 \times 10^{12}$~Hz to over
$10^{18}$~Hz (see 3C 273 or 3C 279 and MKN 501 in Fig. \ref{fig:sed})
reflecting the maximum energy at which particles can be accelerated
\cite[e.g.][]{GiommiPlanck}. Blazars with rest-frame \nup~$< 10^{14}$~Hz
are called Low Synchrotron Peaked (LSP) sources, while those with
$10^{14}$~Hz $<$ \nup~$< 10^{15}$~Hz and \nup~ $> 10^{15}$~Hz are called
Intermediate and High Synchrotron Peaked (ISP and HSP) sources,
respectively \citep{abdosed}. These definitions extend the original division
of BL Lacs into LBL and HBL sources \cite{padgio95}. An expanded version 
of the work presented here is given by \cite{giommi2012}.

\section{CURRENT STATUS}

The two main blazar subclasses have many differences, which include:

\begin {enumerate}

\item {\it different optical spectra} (by definition). There are, however,
  a number of BL Lac - FSRQ transition objects, which include even BL
  Lacertae itself, the prototype of the class, which displays at times
  moderately strong, broad lines \cite[e.g.][]{ver96} and 3C 279, a
  well-studied FSRQ, which can appear nearly featureless in a bright state
  \cite{pia99};

\item {\it different extended radio powers}. Most BL Lacs have extended
  radio powers and morphologies consistent with those of Fanaroff-Riley
  (FR) type I, while basically all FSRQs are FR II-like \cite[][and
    references therein]{UP95}. However, some radio-selected BL
  Lacs are known to posses an FR II-like structure \cite[e.g.][]{rec01};

\item {\it very different redshift distributions.} FSRQs, similarly to
  radio quiet QSOs, are typically found at redshifts $\sim 1-2$, and up to
  $\sim 5.5$, while BL Lacs are usually much closer with very few cases at
  $z \gsim 0.6$ \cite{mas09}. A large fraction of BL Lacs, however,
  \cite[$\sim 43$\% in BZCAT and $> 50 - 60$\% of the BL Lacs in the Fermi
    1 and 2 year AGN catalogs:][]{fermi1lac,fermi2lac} have no measured
  redshift, due to the lack of any detectable feature in their optical
  spectrum, despite the use of 8/10-m class optical telescopes for the
  spectroscopy identification campaign;
     
\item {\it different cosmological evolutions.} Radio and X-ray selected
  samples have shown that the two blazar subclasses have very different
  cosmological evolutions, with FSRQs evolving strongly (again similarly to
  radio quiet QSOs) and BL Lacs evolving at a similar, or perhaps lower
  rate, in the radio band, or even showing no or negative evolution in the
  X-ray band \cite[e.g.][]{stickel1991,rec00,pad07};

\item {\it widely different mix of FSRQs and BL Lacs in radio and X-ray
  selected samples}, with the latter typically including a much larger
  fraction of BL Lacs than the former. In fact, while only $\sim 15\%$ of
  WMAP5 blazars are BL Lacs (Section \ref{ingredients}), this fraction is
  instead $\sim 70\%$ in the EMSS complete sample, which includes 41 BL
  Lacs and 15 FSRQs \cite{rec00,pad03};

\item {\it widely different distributions of the synchrotron peak energy
  \nup.}  The rest-frame \nup~distribution of FSRQs is strongly peaked at
  low energies ($ \langle$ \nup $\rangle=10^{13.1\pm0.1}$ Hz) and never
  reaches very high values (\nup $\lsim 10^{14.5}$ Hz) independently of the
  selection method \cite{GiommiPlanck}, while the \nup~ distribution of BL
  Lacs is shifted to higher values by at least one order of magnitude. It
  can also reach values as high as \nup $\gsim 10^{18}$ Hz and its shape
  varies strongly depending on the selection band \cite[that is radio,
    X-ray or $\gamma$-ray:][]{abdosed,GiommiPlanck}.
  
\end {enumerate}

Some of these differences have been explained by so-called unified schemes,
which posit that BL Lacs and FSRQs are simply FR I and FR II radio galaxies
with their jets forming a small angle with respect to the line of sight
\cite{UP95}. Radio galaxies would then be the ``parent'' population of
blazars. Due to relativistic beaming this has enormous effects on their
apparent emitted power and luminosity functions (LFs) and can explain their
different extended radio powers and, partly, their cosmological
evolutions. However, unified schemes per se cannot account for transition
objects, the different evolution of radio and X-ray selected BL Lacs, 
and \nup~distributions. 

\section{A NEW SCENARIO FOR BLAZARS}

In our new scenario, the observed blazar optical spectrum is the result of
a combination of an intrinsic EW distribution and the effects of three
components: a non-thermal, jet related one, a thermal one due to the
accretion disk, and emission from the host galaxy. Different mixes of these
components determine the appearance of the optical spectrum and therefore
the classification of sources in FSRQs (dominated by strong lines), BL Lacs
(with diluted, weak lines, if a standard accretion disk is present), and
radio-galaxies (where the host galaxy swamps both the thermal and
non-thermal nuclear emission present in blazars). The other novel component
is a single LF whose evolution depends on radio power.

\subsection{Simulation ingredients}\label{ingredients}

Our idea was tested through extensive Monte Carlo simulations, which
include the following ingredients, which we kept as simple as possible and
tied as much as possible to observational data:

\begin{enumerate}

\item{\bf Luminosity function} We derive the LF and evolution of blazars at
  41 GHz from the Wilkinson Microwave Anisotropy Probe (WMAP5) sample
  \cite{wri09}, which is practically equivalent to a radio-selected
  sample. We extend on the work of \cite{GiommiWMAP09} and define a
  flux-limited sample of high Galactic latitude sources ($f_{41GHz} \ge
  0.9$ Jy, $|b_{\rm II}| > 15^{\circ}$) including 161 FSRQs, 29 BL Lacs,
  and 10 blazars of unknown type. By applying a maximum likelihood
  technique to the WMAP5 blazars we obtain, together with the evolution
  discussed below, a best-fit local LF $\Phi(P) \propto P^{-3}$ (in units
  of Gpc$^{-3}$ P$^{-1}$) between $1.9 \times 10^{24}$ and $4.2 \times
  10^{27}$ W/Hz, which we assume in our radio simulations. 
  Throughout this paper we use a $\Lambda$CDM cosmology with $H_0 = 70$ km
s$^{-1}$ Mpc$^{-1}$, $\Omega_m = 0.27$ and $\Omega_\Lambda = 0.73$.
  
\item{\bf Cosmological evolution} Powerful ($P_{\rm r} \gsim 10^{26}$ W/Hz)
  radio sources display strong evolution at low redshifts followed by a
  decline at higher redshifts \cite[e.g.][]{wa05}. We parametrize this
  behavior with a model of the type $P(z) = (1+z)^{k+\beta z}$, which
  allows for a maximum in the luminosity evolution followed by a decline. A
  maximum likelihood technique applied to the WMAP5 sample allows us to
  derive $k = 7.3$ and $\beta = -1.5$ in the $0 - 3.4$ redshift range
  (which implies a peak at $z \sim 1.85$), which we assume in our
  simulations. Lower luminosity ($P_{\rm r} \lsim 10^{26}$ W/Hz), mostly FR
  I radio sources display a much weaker cosmological evolution, which
  reaches $\approx$ zero at $P_{\rm r} < 10^{25}$ W/Hz \cite[e.g.][and
    references therein]{gen10}. We took this into account by using the
  radio LFs of BL Lacs and FSRQs derived from those of FR Is and FR IIs and
  based on the beaming model of \cite{UP95}. We then used the fraction of
  beamed FR I blazars in bins of radio power to simulate the fraction of
  non-evolving radio sources as a function of power.
  
\item{\bf Non-thermal component} To represent the non-thermal/jet
  component, we assume a simple homogeneous synchrotron self-Compton model
  \cite[SSC, see, e.g.][and references therein]{tra09} with relativistic
  electrons distributed as a power law at low energies and as a
  log-parabola at high energies \cite{mas06}. This model represents well
  the synchrotron part of the observed SEDs, which always extends at least
  to the optical band where the classification of blazars as FSRQs or BL
  Lacs occurs. As for the inverse Compton emission, which can be important
  in the soft X-ray band, we set the Compton dominance so as to reproduce
  the observed $f_{\rm x}/f_{\rm r}$ in FSRQs. The Lorentz factors of the
  electrons radiating at the peak of the synchrotron SED component
  ($\gamma_{\rm peak}$) cover the range $\sim 10^{2.5} - 10^{4.5}$, which is that
  expected for typical parameters of the SSC model as shown in Fig. 36 of
  \cite{abdosed}. The shape of the distribution was chosen 
  to  reproduce the observed \nup~distributions in radio and
  X-ray selected samples of blazars. We assumed a mean value of 15 for the
  Doppler factor $\delta$. This was chosen to be consistent with the mean
  superluminal speed $\beta_{\rm app} \sim 12$ obtained by \cite{lis09},
  and with the typical Lorentz factor $\Gamma \sim 15$ derived by
  \cite{hov09} (since for the angle that maximizes the apparent velocity
  $\delta \sim \beta_{\rm app} \sim \Gamma$).
     
\item{\bf Accretion Disk and broad emission lines} We use the quasar
  spectral template of \cite{van01} \cite[see Fig. 4 of][]{GiommiPlanck}. A
  standard accretion disk is likely to be present only in so-called ``high
  excitation'' radio galaxies (HERGs), while it appears not to be there, or
  be less efficient, in low-excitation ones (LERGs). Almost all FR Is are
  LERGs, while most FR IIs are HERGs, although there is a population of FR
  II LERGs as well. Observational evidence suggests that LERGs, which means
  all FR Is and some FR IIs, either do not possess an accretion disk, or if
  the disk is present is much less efficient than in FR IIs (i.e. of the
  Advection Dominated Accretion Flow [ADAF] type) \cite[e.g.][]{ev06} .  We
  have then associated the presence of a standard accretion disk only with
  beamed FR II sources and assumed that all those with an FR I parent (the
  non-evolving sources) have no disk.
  
\item{\bf Equivalent width distributions} The intrinsic (before dilution)
  distributions of the EW of the broad lines (Ly$\alpha$, C~IV, C~III,
  Mg~II, H$\beta$, H$\alpha$) have been assumed to be those of a sub-sample
  of the radio quiet QSOs included in the SDSS DR7
  database (http://www.sdss.org/dr7/).  We assumed Gaussian
  distributions characterized by the measured means ($\langle$EW${_{{\rm
        H}\alpha}}$$\rangle$= 200 \AA, $\langle$EW${_{{\rm
        H}\beta}}$$\rangle$= 23 \AA, $\langle$EW${_{{\rm
        Mg-II}}}$$\rangle$= 18 \AA, $\langle$EW${_{{\rm C-III}}}$$\rangle$=
  16 \AA, $\langle$EW${_{{\rm C-IV}}}$$\rangle$= 20 \AA,
  $\langle$EW${_{{\rm Ly}\alpha}}$$\rangle$= 47 \AA) and dispersions for
  the various lines.
 
\item{\bf The disk to jet power ratio} The disk and jet components in
  blazars are known to be correlated and possibly of the same order of
  magnitude \cite[see, e.g.][]{del03,ghis11}. We are interested in the
  somewhat simpler question of determining how the luminosity of the
  accretion disk (blue bump intensity at 5000 \AA) scales with radio power
  (at 5 GHz). The relevant data were derived by using the very large amount
  of multi-frequency information included in public databases and the tools
  that are now available to analyze SEDs \cite{SEDtool}. Figure
  \ref{fig:sed} gives some examples of representative objects. The amount
  of thermal flux in each FSRQ was estimated by matching the composite
  optical QSO spectrum of \cite{van01} to the SED data in the blazar
  rest-frame.  Fig. \ref{fig:sed} gives examples of the matching of the
  composite QSO spectrum (blue line) to the data for the case of 3C 273 or
  3C 279. An upper limit was estimated for BL Lac objects by placing the
  composite QSO spectrum in the SED at an intensity such that the optical
  lines would not be detectable in the optical spectrum (typically a factor
  ten below the observed flux. Fig. \ref{fig:sed} illustrates the case of
  MKN501). This was done for the blazars selected in two surveys, WMAP5 and
  the EMSS. We found that $\alpha_{r-BlueBump}$ (the slope between the 5
  GHz luminosity and the blue bump intensity at 5000~\AA~and defined by
  $L_{\rm disk} = L_{\rm r} (\nu_{\rm 5000\AA}/\nu_{5GHz})^{-\alpha_{\rm
      r-BlueBump}}$) correlates with radio luminosity, although with a
  large scatter. We then fitted a simple linear relationship between the
  two variables ($\alpha_{r-BlueBump}$ = 0.04*log(L$_{\rm radio}) -0.39$)
  and assumed a Gaussian distribution around it with a dispersion of 0.1.

\item{\bf Host galaxy} We have assumed that the host galaxy of blazars is a
  giant elliptical with fixed absolute magnitude of $M_{\rm R} = -22.9$
  \cite{scarpa2000, sba05}. For the spectral shape we used the galaxy
  template of \cite{man01}, who derived it combining the data from 28 local
  elliptical galaxies observed in the wavelength range $0.12 -
  2.4~\mu$m. Figure \ref{fig:sed} shows this template superposed to the SED
  of four well-known blazars.

\end{enumerate}

\subsection{Simulation steps}

Our Monte Carlo simulations start by drawing a random value for the radio
luminosity and redshift based on the luminosity function and evolution
described above and a value of the Lorentz factor of the electron radiating
at the peak of the synchrotron power ($\gamma_{\rm peak}$) from the assumed
distribution. We then calculated the peak of the synchrotron power in the
source rest frame by assuming a simple SSC model
($\nu_{peak}=3.2\times10^6\gamma_{\rm peak}^2 B \delta$).  The magnetic
field was fixed to B = 0.15 Gauss and the Doppler factor $\delta$ was
randomly drawn from a gaussian distribution with $\langle \delta \rangle
=15$ and $\sigma$= 2. We then calculated the observed radio flux density
(from the radio luminosity and redshift) and the non-thermal emission in
the optical and X-ray bands under the assumption that the spectral shape of
the observed emission is a log parabola around \nup\ and that the low
energy part of the SED (cm and mm wavelengths) is a power law (as is
typically seen in blazars: see e.g. Fig. \ref{fig:sed} for some
representative examples). We then added an accretion (blue bump) component
as described above (only for beamed FR II sources), re-scaling the SDSS
quasar template to this value and drew a random value of the equivalent
width of Ly$\alpha$, C~IV, C~III, Mg~II, H$\beta$, H$\alpha$ starting from
the EW distribution observed in the SDSS radio quiet QSOs. The host galaxy
optical light was added assuming a standard giant elliptical and we then
calculated the total optical light and the observed equivalent width of all
the broad lines considered by taking into account the dilution due to the
non-thermal and host galaxy optical light. Sources were classified as FSRQs
if the rest-frame EW of at least one of the broad lines that enter the
optical band in the observer frame (which we assume to cover the $3,800 -
8,000$ \AA~range) was $> 5$~\AA. Otherwise, the object was classified as a
BL Lac, unless the host galaxy dominated the optical light causing the Ca
H\&K break to be larger than 0.4 \cite{mar96,lan02}, in which case the
source was classified as a radio galaxy. A BL Lac whose maximum EW is $< 2$
\AA,\, or for which the non-thermal light was at least a factor 10 larger
then that of the host galaxy \citep{piranomonte07}, was deemed to have a
redshift which cannot be typically measured.
  
It is important to stress that the scope of our simulations is {\it not} to
reproduce {\it all} the observational details. While that could be possible
in theory, in practice it would require a large number of parameters and
some speculations. Our approach is instead to keep the number of
assumptions to a minimum, to obtain robust,
almost model-independent conclusions.

We are confident that our main results are stable to changes in input
parameters.  The adopted LF and evolution were in fact varied by $1 \sigma$
from the WMAP5 best fit and we also used as LF the sum of the BL Lac and
FSRQ LFs based on the beaming model of \cite{UP95}. We also run the
simulations with values of $\langle\delta\rangle$ in the 5 to 20 range, and
considering also a dependence of $\delta$ on radio power
\citep[e.g.][]{hov09}. No major changes were obtained as compared to our
default assumptions.

\subsection{Simulations of radio and X-ray surveys}

We simulated a radio flux density limited survey with $f \ge 0.9$ Jy, to
match the WMAP5 sample, and an X-ray flux limited survey down to $5\times
10^{-13}$ \ergs in the $0.3 - 3.5$ keV band, in order to be able to compare
it with the EMSS. To ensure good statistics each simulation run included
10,000 sources.  In the X-ray case, since radio powers reach lower values
than in the radio case, we extrapolated the radio LF down to $1.9 \times
10^{23}$ W/Hz assuming the same slope.

\section{COMPARING SIMULATIONS AND REAL DATA}

\subsection{ Radio flux density limited survey}\label{radio_survey}

\begin{table}[t]
\begin{center}
\caption{Results from a simulation of a radio flux density limited survey (0.9 Jy)}
\begin{tabular}{llcc}
\hline \textbf{Source type} & \textbf{Number} & \textbf{$\langle z \rangle$} & \textbf{$\langle V/V_{\rm m} \rangle$} 
\\
FSRQs      &    ~7,587        &  1.24 & 0.64  \\
BL Lacs     &     ~1,879 (1,191)   &  0.87 & 0.60 \\
Radio galaxies    &     ~~~534     &  0.04 & 0.48   \\
\hline Total     &    10,000       &  1.13 & 0.63  \\
\hline
\end{tabular}
\label{tab:radiosim}
\end{center}
\end{table}

\begin{figure}
\includegraphics[height=8.cm,angle=-90]{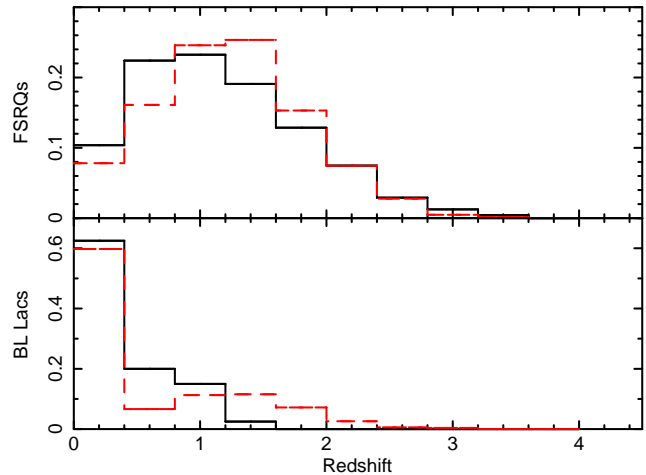}
\vspace{0.8cm}
\caption{Top panel: the redshift distribution of the WMAP5 FSRQs (solid
  histogram) compared to that of FSRQs in a simulation of a radio survey
  (dashed histogram). Bottom panel: the redshift distribution of the WMAP5
  BL Lacs (solid histogram) compared to that of BL Lacs in a simulation of
  a radio survey (dashed histogram)}
 \label{fig:wmap_red}
\end{figure}

\begin{figure}
\includegraphics[height=8.cm,angle=-90]{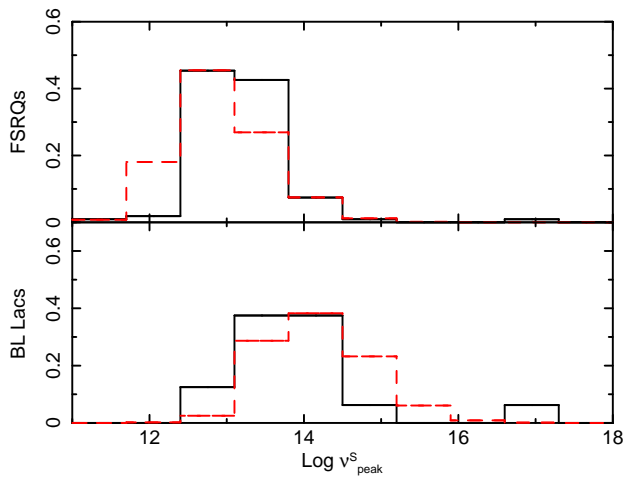}
\vspace{0.5cm}
\caption{Top panel: the \nup~distribution of radio selected FSRQs taken
  from the work of \cite{GiommiPlanck} (solid histogram) compared to that
  of FSRQs in a simulation of a radio survey (dashed histogram). Bottom
  panel: the \nup~distribution of the BL Lacs in the radio sample of
  \cite{GiommiPlanck} (solid histogram) compared to that of BL Lacs in a
  simulation of a radio survey (dashed histogram).}
 \label{fig:radio_nupeak}
\end{figure}

Table \ref{tab:radiosim} summarizes our main results by giving the number
of sources per class, their mean redshift, and \avvovm~(where $V$ is the
volume out to the source and $V_{\rm m}$ is the volume at the distance
where the object would be at the flux limit: \cite{sch68}). The number in
parenthesis refers to the BL Lacs with measurable redshift, to which the
mean redshift and \avvovm~pertain. About 3/4 of our sources are classified
as FSRQs, with the fraction of BL Lacs being $\sim 19.8\%$ of blazars,
which is consistent with the value of $15.3^{+3.7}_{-3.0} \%$ in the WMAP5
sample. A small fraction ($5 \%$) of the simulated blazars are classified
as radio galaxies. These are bona-fide blazars misclassified because their
non-thermal radiation is not strong enough to dilute the host galaxy
component.  The mean redshift for our simulated FSRQs agrees with the WMAP5
value of $1.13$, while for BL Lacs this is slightly larger than the WMAP5
mean ($0.55$). Fig. \ref{fig:wmap_red} shows the overall good agreement
between our simulated redshift distributions (where we have only included
sources with a measurable redshift) and the observed ones. $63\%$ of our BL
Lacs have a redshift determination, in excellent agreement with the WMAP5
value of $69^{+27}_{-20} \%$.  $79\%$ of the BL Lacs ($68\%$ of those with
redshift) have a standard accretion disk and are therefore broad-lined but
are classified as BL Lacs only because their observable emission lines are
swamped by the non-thermal continuum.
As regards \avvovm, our simulated mean values agree with the observed
ones of $0.62\pm0.02$ and $0.63\pm0.05$ for WMAP5 FSRQs and BL Lacs
respectively \citep[cf. also the value of $0.60\pm0.05$ for the 1 Jy BL Lac
sample:][]{stickel1991}.
Fig. \ref{fig:radio_nupeak} compares the distributions of \nup, the
synchrotron peak energy, of sources classified as FSRQs and BL Lacs in our
simulation with those of blazars included the radio sample of
\cite{GiommiPlanck}, which is the sample with the best determination of
\nup~values currently available. The agreement is clearly quite good and
reproduces well the fact that BL Lacs tend to have \nup~values
significantly higher than FSRQs.

\subsection{X-ray flux limited survey}

\begin{table}[t]
\begin{center}
\caption{Results from a simulation of an X-ray  flux limited survey ($5\times 10^{-13}$ \ergs)}
\begin{tabular}{llcc}
\hline \textbf{Source type} & \textbf{Number} & \textbf{$\langle z \rangle$} & \textbf{$\langle V/V_{\rm m} \rangle$} 
\\
FSRQs      &    ~2,836       &  1.23 & 0.65  \\
BL Lacs (all)   &     ~5,622 (4,460)    &  0.36 &  0.51 \\
BL Lacs  (log \nup~$> 16.5$) &     ~~~927 (895)      &  0.33 &  0.45 \\
BL Lacs (log \nup~$> 17$) &     ~~~185 (177)      &  0.34 &  0.34 \\
Radio galaxies    &    ~1,542    &  0.04 & 0.48   \\
\hline Total     &    10,000        &  0.58 &   0.55  \\
\hline
\end{tabular}
\label{tab:xsim}
\end{center}
\end{table}

Table \ref{tab:xsim} summarizes our main results. About 2/3 of our blazars
are classified as BL Lacs, which is consistent with the value of
$73^{+19}_{-15} \%$ in the EMSS sample. As in the radio case, a small
fraction ($15\%$) of the simulated blazars are misclassified as radio
galaxies. The mean redshifts for our simulated FSRQs and BL Lacs are in
reasonable agreement with the EMSS blazar sample values $\sim 1$ and $\sim
0.37$. We note that, unlike the WMAP5 sample, the EMSS sample is relatively
small (56 sources) and therefore a detailed comparison is hampered by the
small number statistics. $79\%$ of our BL Lacs have a redshift
determination, in good agreement with the EMSS value of $93^{+26}_{-21}
\%$. Although we assumed that all non-evolving sources do not have a
standard accretion disk, $30\%$ of the BL Lacs possess one and are
classified as BL Lacs only because their emission lines are swamped by the
non-thermal continuum. The smaller fraction of X-ray selected BL Lacs with
disks in our simulations, as compared to radio-selected ones, is in
accordance with the fact that fewer EMSS BL Lacs have emission lines
clearly detectable in their optical spectra than, for example, 1 Jy BL Lacs
\cite{rec00,rec01,stickel1993}. As regards  \avvovm, our simulated
mean values agree reasonably well with the EMSS ones of $0.67\pm0.08$ and
$0.42\pm0.05$ for FSRQs and BL Lacs respectively, derived using the samples
described in \cite{pad03}.
\begin{figure}
\includegraphics[height=8.cm,angle=-90]{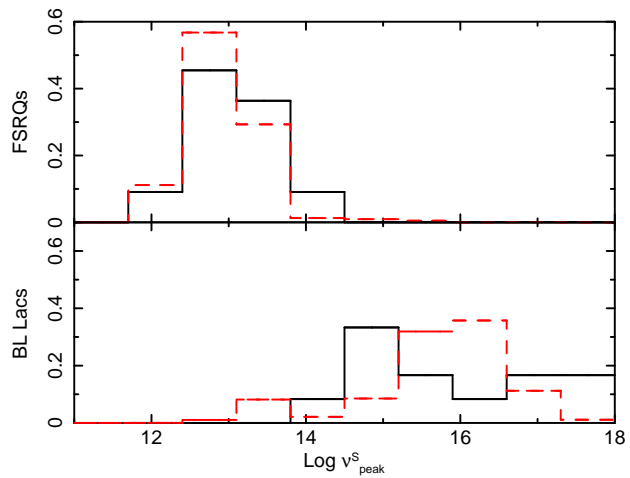}
\vspace{0.5cm}
\caption{Top panel: the \nup~distribution of the X-ray selected FSRQs in
  \cite{GiommiPlanck} (solid histogram) compared to that of FSRQs in a
  simulation of an X-ray survey (dashed histogram). Bottom panel: the
  \nup~distribution of the BL Lacs in the X-ray flux limited sample of
  \cite{GiommiPlanck} (solid histogram) compared to that of BL Lacs in a
  simulation of an X-ray survey (dashed histogram).}
\label{fig:x_nupeak}
\end{figure}
Fig. \ref{fig:x_nupeak} compares the distributions of \nup~of FSRQs and BL
Lacs in our simulation with those of blazars belonging to the soft X-ray
sample of \cite{GiommiPlanck}, which includes \planck, \swift and \fermi
observed blazars and it is therefore probably the sample with the best
determination of \nup~values currently available. Our simulations reproduce
the fact that BL Lacs have much higher \nup~values than FSRQs.

\subsection{A blazar sequence?}\label{sec:seq}

\begin{figure*}
\includegraphics[height=12.cm,angle=-90]{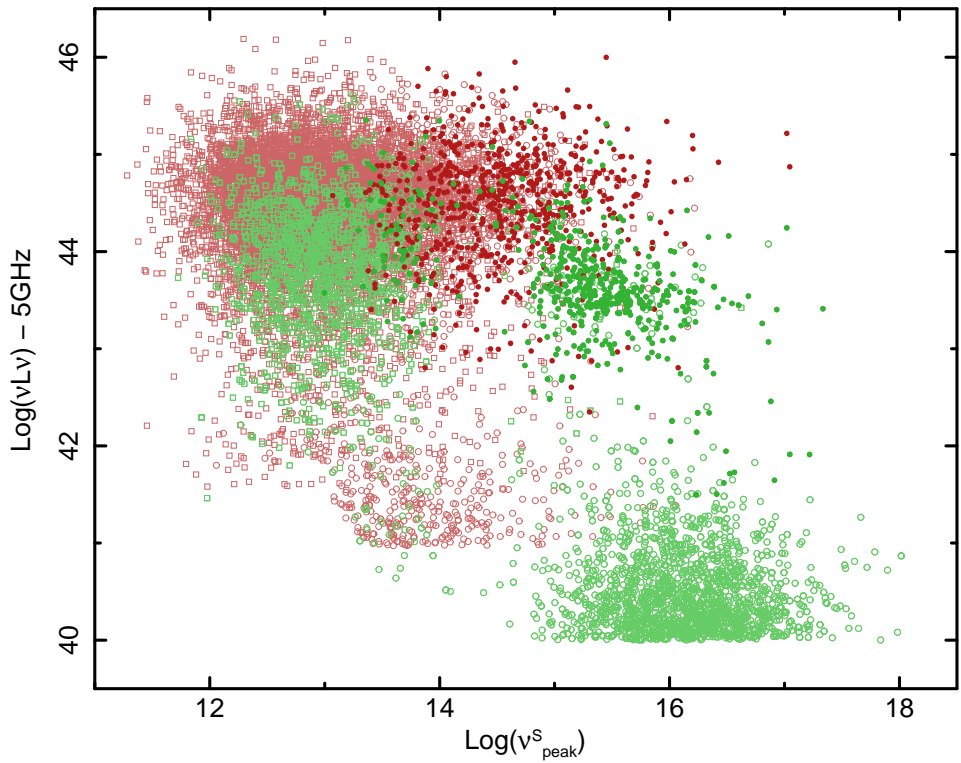}
\vspace{0.9cm}
\caption{The FSRQs (open squares) and BL Lacs (circles) of our radio flux
  density limited ($f_{\rm r} > 0.9$ Jy, red symbols) and X-ray flux
  limited ($f_{\rm x} > 5\times 10^{-13}$\ergs , green symbols) simulated
  samples, plotted in the log($\nu$)-log($\nu$L$_{\nu}$) plane, the portion
  of the parameter space used by \cite{fossati98} to show the existence of
  the ``blazar sequence'' by comparing radio ($f_{\rm r} > 1-2$ Jy) and
  X-ray selected ($f_{\rm x} \gsim 10^{-12}$\ergs) complete samples.
  Filled circles represent BL Lacs with very weak lines (EW $< 2~$\AA) or
  completely featureless which, in a real survey would most probably not
  have a measured redshift, and therefore would not appear in the plot.}
\label{fig:sequence}
\end{figure*}

The existence of an anti-correlation between bolometric luminosity and
\nup, the ``blazar sequence", has been discussed since first proposed by
\cite{fossati98} and \cite{ghis98} \citep[e.g.,][]{giommi_sedent_I,pad03,
  padovani07,ghisellinitavecchio08,nieppola08,GiommiPlanck}.  Figure
\ref{fig:sequence} shows our radio and X-ray selected simulated blazars in
the log(\nup) -- log($\nu$L$_{\nu}$(5 GHz)) plane. This reproduces the plot
used by \cite{fossati98} to propose the existence of the blazar sequence
based on the correlation shown in this plane by FSRQs and BL Lacs
discovered in shallow radio surveys (2 and 1 Jy samples) and BL Lacs found
in the X-ray flux limited {\it Einstein} slew survey ($f_{\rm x} \gsim
10^{-12}$\ergs). Indeed, considered together the simulated radio and X-ray
selected blazars display a broad correlation with radio selected FSRQs and
BL Lacs (red open squares and open circles) mostly filling the top left and
central part of the diagram and X-ray selected BL Lacs (green open circles)
mostly confined to the lower right corner of the plot. This particular
positioning of the points (bright FSRQs of the LSP type vs. fainter HSP BL
Lacs) is not due to any intrinsic correlation between luminosity and
\nup~but results from the fact that bright radio sources are mostly drawn
from the high end of the blazar luminosity function, while BL Lacs in X-ray
flux limited samples are mostly high \nup~sources (intrinsically rare)
drawn from the low end of the luminosity function where the source density
is largest. The most important difference between Fig. \ref{fig:sequence}
and the diagram of \cite{fossati98} is in the high-luminosity - high
\nup~part, where most of the radio and X-ray selected objects with no
redshift (red and green filled points) are located. These sources could not
be plotted by \cite{fossati98} since the luminosity of blazars without
redshift cannot be estimated. This made the top right part of the
\cite{fossati98} diagram empty, thus contributing to making the data points look
like a sequence.


\section{DISCUSSION}\label{discussion}

\subsection{Evolution}

A long-standing blazar puzzle is the difference in redshift distribution
and cosmological evolution between FSRQs and BL Lacs, selected both in the
radio and in the X-ray band. BL Lacs are mostly located at low redshifts
and exhibit moderate, or even negative, evolution, while FSRQs evolve
strongly just like radio quiet QSOs and show a redshift distribution that
peaks at $z > 1$ \cite{stickel1991,rec00,beck03,pad07,GiommiWMAP09}.  Our
simulations reproduce quite well both of these findings (see
Fig. \ref{fig:wmap_red} and Tables \ref{tab:radiosim} and \ref{tab:xsim})
implying that they are due to heavy selection effects.  Most of the
simulated BL Lacs found in radio surveys are luminous objects with broad
lines that are diluted by non-thermal radiation beyond the 5~\AA~EW limit
(many of them just below, thus allowing a measurement of their redshift),
while the BL Lacs found in simulated X-ray surveys typically show high
\nup~values (and therefore are X-ray bright) and are drawn from the
low-power end of the radio luminosity function where non-evolving FR Is are
preferentially found (Section \ref{ingredients}). We note that the
\avvovm~of radio selected BL Lacs is not too different from that of FSRQs,
while the \avvovm~of X-ray selected BL Lacs is significantly lower, as
found in real surveys (Section \ref{radio_survey}).  Another interesting
outcome of our simulations is the fact that X-ray selected BL Lacs with
progressively larger values of \nup~are characterized by lower and lower
values of \avvovm~(see Tab. \ref{tab:xsim}). This is in full agreement
with the puzzling, and so far unexplained, results of \cite{giommi_sedent_I} 
and \cite{rec00} who reported that the \avvovm~of BL Lacs is a
function of their X-ray-to-radio flux ratio (which in turn depends on \nup).

\subsection{\nup~distribution}

Recent results, based on radio and $\gamma$-ray surveys, have revealed that
BL Lacs, on average, display a distribution of \nup~energies, which is
shifted to values higher than those of FSRQs \citep{abdosed,GiommiPlanck},
expanding on the well-known fact that high \nup~objects (HSPs) are always
BL Lacs.  This experimental difference is well reproduced in our
simulations (see Fig. \ref {fig:radio_nupeak}) which give $\langle
$log(\nup)$\rangle$ = 12.9 for FSRQs and $\langle $log(\nup)$\rangle$ =
14.1 for BL Lacs for the case of a radio survey. This distinction is due to
the fact that blazars with higher \nup~values produce more non-thermal
optical light than low \nup~sources, diluting more easily the broad line
component, and are therefore classified more frequently as BL Lac
objects. We note that this has been interpreted in the literature as an
intrinsic physical difference between LSP (detected mostly in the radio
band) and HSP (detected mostly in the X-ray and \gr band) blazars due to
the fact that HSPs are observationally characterized by a low intrinsic
power and external radiation field, given their very weak or absent
emission lines. As a consequence, cooling was thought to be less dramatic
in HSP allowing particles to reach energies high enough to produce
synchrotron emission well into the X-ray band \citep{ghis98}. In our
scenario, instead, all sources have exactly the same chance of being HSP or
LSP (that is, the value of $\gamma_{\rm peak}$ is drawn from a distribution
independently of luminosity) and the very different \nup~distributions
observed in radio and X-ray surveys arise from the strong selection effect
discussed in Section \ref{sec:seq} and the emission line dilution mentioned
above.
\begin{figure}
\includegraphics[height=7.4cm,angle=-90]{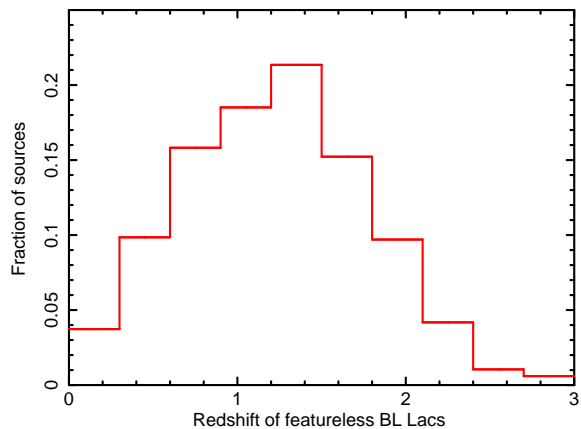}
\vspace{0.8cm}
\caption{The redshift distribution of the BL Lacs that show a featureless
  spectrum in our simulation of a radio flux density limited survey and
  that in a real survey would have no redshift determination.}
 \label{fig:zdistOfnoredshiftSources}
\end{figure}
Our simulations predict the existence of a significant number of BL Lacs
with redshift that cannot be measured, which can occur when both \nup~and
radio power are so large that dilution becomes extreme. For example, $\sim
81\%$ of our simulated sources with \nup~$> 10^{15}$ Hz and $P_{\rm r} >
10^{26}$ W/Hz have no redshift. This is consistent with the fact that most
BL Lacs in current \gr~selected samples have no measured redshift, as
\fermi is known to preferentially select high \nup~BL Lacs
\citep{fermi1lac,fermi2lac}. This effect is also clearly shown in
Fig. \ref{fig:sequence} where most of the simulated blazars with no
measurable redshift (filled circles with light colors) occupy the top right
part of the diagram.  Fig. \ref{fig:zdistOfnoredshiftSources} shows the
intrinsic redshift distribution of these featureless BL Lacs for the case
of our simulation of a radio flux density limited survey.

\subsection{What is a BL Lac?}

It was originally suggested that the absence of broad lines in BL Lacs was
due to a very strong, Doppler-boosted synchrotron continuum
\cite{bla78}. In the years following that paper observations of various BL
Lacs, mostly selected in the X-ray band, showed that in many cases their
optical spectrum was not swamped by a non-thermal component, as host galaxy
features were very visible, and it was thought that most BL Lacs had
intrinsically weak lines \cite{sto91}. We have shown here that these two
possibilities are not mutually exclusive and indeed are both viable,
depending on radio power and, therefore, on the band of selection. One
important consequence of our scenario is that objects so far classified as
BL Lacs on the basis of their {\it observed} weak, or undetectable,
emission lines belong to two physically different classes: intrinsically
weak-lined objects, whose parents are LERGs/FR Is (more common in X-ray
selected samples, since they reach lower radio powers) and heavily diluted
broad-lined sources, which are beamed HERGs/FR IIs (more frequent in radio
selected samples). Therefore, while the non-thermal engine is probably the
same, the thermal one is obviously different. This solves at once the issue
of the FSRQ/BL Lac transition objects and of the many differences between
BL Lacs selected in the radio and X-ray bands, which include line strength,
extended radio emission and morphology, and evolution \citep[e.g.][and
  references therein]{rec01}. It also implies that BL Lacertae, the
prototype of the class, is actually {\it not} a BL Lac but an FSRQ with its
emission lines swamped by the jet. This hypothesis, which explains also
many other open issues of blazar research, is testable. Our simulations, in
fact, imply that the majority of sources in high-flux density
radio-selected samples are identified as BL Lacs only because all lines
have EW~$<~5$~\AA~{\it in the optical observing window}. These sources
should show a strong (EW $> 5$ \AA) H$\alpha$ line in the near or
mid-infrared, since this is the strongest emission line, and would then be
considered FSRQs. HERG/FR II BL Lacs, then, should be easily recognizable
through infrared spectroscopy. These sources, as expected, are also more
dominant at higher radio powers: for example, according to our radio
simulation the fraction of BL Lacs with standard accretion disks is only
$\sim 3\%$ for $P_{\rm r} \le 10^{26}$ W/Hz but becomes $\sim 98\%$ above
this value. This also means that fainter radio-selected samples of BL Lacs
should be more and more similar to X-ray selected ones, apart from their
\nup~values, since higher \nup~are easier to detect in the X-rays
\citep{padgio95}.  The implications of our hypothesis for unified schemes
is quite straightforward: the parent population of BL Lacs need to include
both LERGs/FR Is and HERGs/FR IIs, while that of FSRQs is made up of
HERGs/FR IIs only. This should have only a small effect, for example, on
the LF fitting done by \cite{UP95}, as HERG/FR II BL Lacs make up the
high-power end of the radio LF while most of the number density comes from
LERG/FR I BL Lacs.

\subsection{Blazar classification}

Our new scenario has strong implications on blazar classification.  If the
relevant physical distinction for radio sources is between LERGs (mostly FR
Is) and HERGs (FR IIs), for most purposes then (LFs, evolution, etc.)
HERG/FR II BL Lacs should be simply grouped with FSRQs. How does one
distinguish in practice HERG/FR II BL Lacs from LERG/FR I BL Lacs?  This is
simple in the presence of {\it any} (even weak) broad lines or for
transition objects. In other cases (e.g. completely featureless spectrum or
in presence of absorption features) there is no easy way to distinguish
between the two subclasses although, for example, radio power and/or
morphology could help. Given the paucity of known LERG/FR Is at relative
high redshifts, X-ray selected (and also fainter radio-selected) BL Lac
samples are also useful in selecting such sources, which are very relevant
also for the study of the so-called ``AGN feedback'' and the role that AGN
radio emission plays in galaxy evolution through the so-called
``radio-mode'' accretion \citep{cro06}. It should also be clear that BL
Lacs can be used to study the broader issue of the relationship between
LERGs and HERGs, including their evolution. A by-product of our simulations
has also been the realization that some sources classified as
radio-galaxies do {\it not} have their jets oriented at large angles with
respect to the line of sight, as expected, but are instead moderately
beamed blazars with their non-thermal emission swamped by the galaxy (note
that none of these objects has a standard accretion disk). These sources,
which are all local ($\sim 90\%$ at $z \le 0.07$) should be recognizable by
their blazar-like SEDs and indeed some of them have already been identified
by \cite{den00}, \cite{giommi_sedent_II} and \cite{ant05}. Recently, a new
classification scheme has been proposed to divide BL Lacs from FSRQs, which
is based on the broad line region (BLR) luminosity in Eddington units and
set at a dividing value of $L_{\rm BLR}/L_{\rm Edd} \sim 5 \times 10^{-4}$
\cite{ghis11}. This turns out to be also the value, which separates
radiatively efficient (i.e., standard accretion disks) from radiatively
inefficient (i.e., ADAFs) regimes, and therefore coincides with our HERG/FR
II -- LERG/FR I division. Therefore, \cite{ghis11} are also suggesting that
HERG/FR II BL Lacs belong with the FSRQs.

\section{TESTS AND PROSPECTS}

Very recently, redshift constraints for 103 blazars from the {\it Fermi}
2LAC catalogue \citep{fermi2lac} have been derived by fitting SED templates
to their UV-to-near-IR multi-band photometry obtained quasi-simultaneously
with {\it Swift}/UVOT and GROND \cite{rau12}. This was done using the
attenuation due to neutral hydrogen along the line of sight at the Lyman
limit to estimate the redshift of the absorber. Eleven of these objects
have $z_{\rm phot} > 1.2$. Some of us have studied the SEDs of these
sources using quasi-simultaneous near-IR to X-ray data. Four blazars turned
out to be of a type never seen before but with properties we were expecting
for some of the BL Lacs without redshift: large \nup~($\sim 5 \times
10^{15}$ Hz) and high-power (see filled circles in Fig. \ref{fig:sequence})
\cite{pad12}. Given their featureless optical spectra, these sources are
therefore most likely high-redshift FSRQs with their emission lines swamped
by the jet, as predicted by our hypothesis.

In this paper we limited our simulations to the radio and the X-ray bands
where SSC is a fair approximation of the observed non-thermal emission. The
properties of \gr~detected blazars are instead not consistent with simple
SSC models \citep[e.g.][]{abdosed}, and almost half of the radio and X-ray
selected LSP blazars are \gr quiet \citep[e.g.][]{GiommiPlanck}. The
present approach must therefore be integrated with additional information
about the properties of the inverse Compton emission before it can be used
to simulate \gr surveys. We are planning to extend our simulations to the
\gr band by taking into account the recent results of \cite{GiommiPlanck}
who determined the \gr properties of blazar samples selected in different
bands. 

\bigskip 
\begin{acknowledgments}
We thank Matteo Perri for providing part of the software used for our Monte
Carlo simulations. We acknowledge the use of data and software facilities
from the ASI Science Data Center (ASDC), managed by the Italian Space
Agency (ASI). Part of this work is based on archival data and on
bibliographic information obtained from the NASA/IPAC Extragalactic
Database (NED) and from the Astrophysics Data System (ADS).
\end{acknowledgments}

\bigskip 

\end{document}